# DeepQuality: Mass Spectra Quality Assessment via Compressed Sensing and Deep Learning

Chunwei Ma, BGI-Shenzhen
Email: machunwei@genomics.cn

ABSTRACT

**Motivation:** Mass spectrometry-based proteomics is among the most commonly used methods for scrutinizing proteomic profiles in different organs for biological or medical researches. All the proteomic analyses including peptide/protein identification and quantification, differential expression analysis, biomarker discovery and so on are all based on the matching of mass spectra with peptide sequences, which is significantly influenced by the quality of the spectra, such as the peak numbers, noisy peaks, signal-to-noise ratios, etc. Hence, it is crucial to assess the quality of the spectra in order for filtering and/or post-processing after identification. The handcrafted features representing spectra quality, however, need human expertise to design and are difficult to optimize, and thus the existing assessing algorithms are still lacking in accuracy. Thus, there is a critical need for the robust and adaptive algorithm for mass spectra quality assessment.
**Results:** We have developed a novel mass spectrum assessment software DeepQuality, based on the state-of-the-art compressed sensing and deep learning algorithms. We evaluated the algorithm on two publicly available tandem MS data sets, resulting in the AUC of 0.96 and 0.92, respectively, a significant improvement compared with the AUC of 0.85 and 0.91 of the existing method SpectrumQuality v2.0.
**Availability:** Software available at https://github.com/horsepurve/DeepQuality

## 1 INTRODUCTION

Mass Spectrometry is a powerful tool for analyzing the ensemble of proteins in cells or organs under different circumstances, to gain insight into the functionalities of proteins. The only data we get from the mass spectrometry instruments are the raw mass spectra and other corresponding information such as retention times, and all the downstream proteomics analyses are based on processing of these spectra. Consequently, the qualities of the spectra are non-ignorable for the success of subsequent analyses. However, in light of the complexity of the signals, even the experienced experts can hardly draw a distinction between so-called "good" and "bad" spectra, i.e. spectra that can be identified or not. Furthermore, the existence of post-translational modification and single amino acid variants makes it much more difficult for a search engine to match a spectrum with its corresponding peptide correctly. As a result, it is crucial to have a method that can assess the quality of a spectrum in supervised or unsupervised manners.

In 2006, Flikka et al. proposed a classification algorithm based on support vector machine (SVM) utilizing 17 manually specified attributes (Flikka *et al.*, 2006), resulting in the area under the ROC curve (AUC) range between 0.73 and 0.91 on various datasets. Similar approaches were reported in

(Salmi *et al.*, 2006) and (Bern *et al.*, 2004). However, these assessment algorithms are still lacking in accuracy, due to the deficiency in feature design. Recent year, deep neural networks have made groundbreaking progress on image recognition and many other tasks (LeCun *et al.*, 2015), making it possible to perform end-to-end assessment of mass spectra, which can be deemed as one-dimensional images.

In this paper, we present an automatic learning process for mass spectra quality assessment. In consideration of the large-scale, sparse nature of mass spectra data, we reduced the dimensionality by means of compressed sensing, and validated its feasibility by robust signal reconstruction. After acquiring the features i.e. compressed signals, we used deep convolutional neural networks for model training. While evaluating on the test data, we got the AUC of 0.96 and 0.92, a significant improvement compared with the existing methods, which validated the credibility of our proposed method.

## 2 METERIALS AND METHODS

### 2.1 Compressed sensing for mass spectra data

Under the assumption of the sparse nature of a signal, Compressed Sensing (CS) can recover it from far few samples than acquired by the Nyquist rate (Candes *et al.*, 2006). Given that most mass spectra are sparse, i.e. peaks will appear in few m/z values compared to all possible m/z values, they can be measured with CS in order for both data compressing and feature extraction (Eleyan *et al.*, 2014).

Formally, for a spectrum containing a set of peaks $s = \{(m_i, I_i)_j\}_{j=1}^{m_s}$, where $m_i$ and $I_i$ are the mass to charge ratio and the intensity for the $j^{th}$ peak in s, we scatter the $m_s$ m/z-intensity pairs into pre-defined bins to form a vector **x** in which every element is the maximum logarithmic-transformed peak intensity within that m/z bin. Then we can use a sensing matrix $A: \mathbb{R}^n \to \mathbb{R}^m$ to perform the sampling, where n and m are the original and reduced dimensions respectively:

$$\mathbf{y} = A\mathbf{x}$$

and the sampled signal **y** serves as the features to be fed into the machine learning algorithm, substitutes for the hand-crafted features used in other literatures.

### 2.2 Deep convolutional network model

After compressed sensing, we used deep convolutional neural network (CNN) (Krizhevsky *et al.*, 2012) for spectra quality classification. As the extracted signals are 1-dimensional vectors, the filters are all of 1-dimension. By using sensitivity analysis, we determined our final network structure as three convolutional layers and in each layer there were 40 filters with filter size of 10. Beside, a fully connected layer with 400 hidden neurons was appended. The activation function of all layers is the rectified linear unit (ReLU) as:

$$ReLU(x) = \begin{cases} x, & x \geq 0 \\ 0, & x < 0 \end{cases}$$

where $x$ is the input to a certain neuron. To avoid overfitting, we applied a dropout ratio of 50% between each layers.

### 2.3 Evaluation

To evaluate the performance of our algorithm, two datasets, "Q-TOF" dataset and "IT" dataset were used. In Q-TOF dataset, the number of "good" and "bad" spectra are 1683 and 8372, while in IT dataset, there are 626 and 7360 "good" and "bad" spectra. The definition of "good" and "bad" spectra follows that of (Flikka *et al.*, 2006). We use five-fold cross-validation to test the performance and the DeepQuality software was run on a NVIDIA Tesla M2070 GPU.

## 3 RESULTS AND CONCLUSION

The pipeline of DeepQuality algorithm is shown in Figure 1. On the two dataset, DeepQuality achieved the AUC (area under the ROC curve) of 0.96 (Q-TOF) and 0.92 (IT) respectively, significantly outperformed those reported in (Flikka *et al.*, 2006) as 0.85 and 0.91, showing the superior ability of compressed sensing in feature extracting and of CNN in accurate classification.

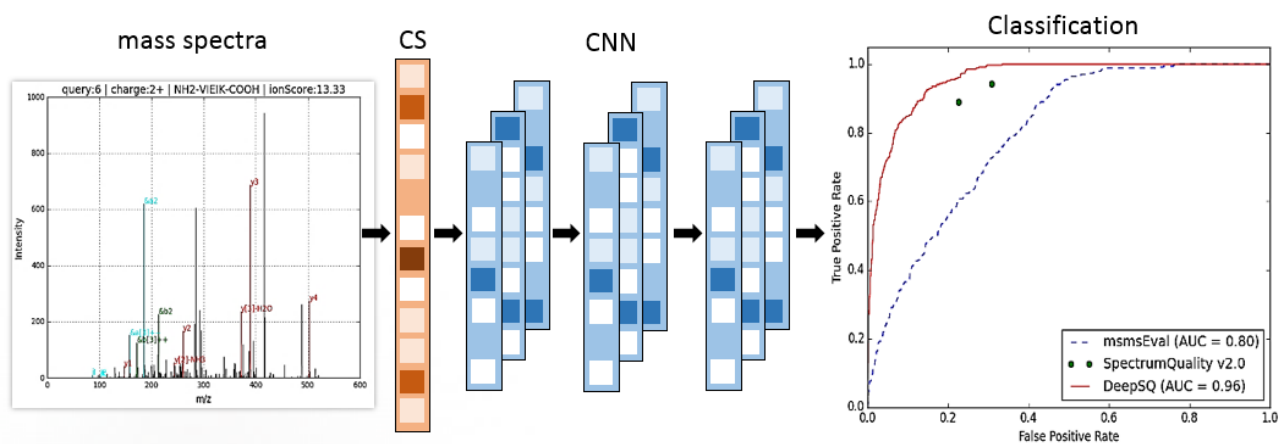

**Figure 1**. The pipeline of DeepQuality. The mass spectra were transformed to feature vectors using compressed sensing (CS), then deep convolutional networks classified them as "good" and "bad". The rightmost panel is the Receiver Operating Characteristic (ROC) curve of our method DeepQuality and msmsEval. Because SpectrumQuality software can only label the spectra as "good" or "bad" without giving a probabilistic score, we plot its result as discrete points (green points).

In compressed sensing theory, the original signal can be re-constructed perfectly using $\ell_1$ minimization, and hence, we hypothesized that the random sample could be used as features, and our result shown their advantage over the handcrafted features. By virtue of the modern deep learning technique, DeepQuality can distinguish between the spectra of high and low qualities and can be used to eliminate the low quality spectra prior to database searching in order to reduce the false discovery rate (FDR) of peptide identification or to recover the unidentified, high quality spectra for further manual analysis. The DeepQuality code and software are freely available at https://github.com/horsepurve/DeepQuality.